# Self-directed growth of AlGaAs core-shell nanowires for visible light applications

Chen Chen[1], Shyemaa Shehata[2], Cécile Fradin[2], Ray LaPierre[1*]

Christophe Couteau[3], Gregor Weihs[3]

[1]Centre for Emerging Device Technologies, Department of Engineering Physics,
McMaster University, Hamilton, Ontario, L8S 4L7, Canada

[2]Department of Physics and Astronomy,
and Department of Biochemistry and Biomedical Sciences
McMaster University, Hamilton, Ontario, L8S 4M1, Canada

[3]Institute for Quantum Computing, University of Waterloo
200 University Avenue West, Waterloo, Ontario, N2L 3G1, Canada

*Corresponding author. Email: lapierr@mcmaster.ca

## Abstract

Be-doped $Al_{0.37}Ga_{0.63}As$ nanowires (NWs) were grown in a molecular beam epitaxy system on GaAs (111)B substrates. Visible light emission was observed at room temperature in a confocal microscope and by micro-photoluminescence (μPL) measurements. μPL polarization measurements at 10 K, and diffusion measurements of NWs in solution, indicated emission from single NWs. Energy dispersive X-ray spectroscopy indicated a core-shell structure and Al composition gradient along the NW axis, producing a potential minimum for carrier confinement. The core-shell structure formed during growth as a consequence of the different Al and Ga adatom diffusion lengths.

Inorganic dots, tubes and wires with nanometer dimensions exhibit electrical and optical properties that may be tuned by their size and shape.[1-3] Among these structures, semiconductor nanowires (NWs) are becoming increasingly important as building blocks in diverse applications including bipolar junction transistors and logic gates, single-electron memory devices, p-n diodes, double barrier resonant tunneling diodes,[4] and biological imaging.[5] For the most part, these NW applications have been realized only in simple material systems such as Si, or binary III-V semiconductors such as GaAs and InP. Reports of NWs containing multiple group-III (e.g., Al, In, Ga) and/or group-V elements (P, As, Sb), of interest in bandgap engineered optoelectronic devices, are comparatively few. Among this class of NWs, the $Al_xGa_{1-x}As$ material system is important for anticipated applications in bio-imaging, displays, solid-state lighting, and single photon sources due to the ability to adjust the bandgap wavelength in the visible spectrum. In this paper, we report growth-related aspects, and demonstrate visible light emission, from AlGaAs NWs where a core-shell structure forms naturally as a consequence of adatom diffusion.

Nanowires were grown by the vapor-liquid-solid (VLS) method in which Au seed particles were used for site selective growth. Substrates of GaAs (111)B were first submitted to a 20-minute uv-ozone treatment, etched in a 10% buffered HF solution, and rinsed with deionized water. The samples were then transported in ambient air to an e-beam evaporation system, where a 1 nm thick film of Au was deposited. The samples with Au were then transferred in ambient air to a gas source molecular beam epitaxy (GS-MBE) growth chamber. In the GS-MBE system, group III species (Al and Ga) were supplied as monomers from a heated solid elemental source, and the group V species were supplied as dimers ($As_2$) from a dual-filament hydride ($AsH_3$) gas cracker operating at 950 ºC. Before the actual growth, the Au-covered substrates were heated to a temperature of 500 °C for 5 minutes under an $As_2$ flux to form Au nanoparticles on the surface. Simultaneous desorption of native oxide was enhanced by the use of an inductively-coupled hydrogen plasma source. After oxide removal and Au nanoparticle formation, the temperature was set to 570 °C for NW growth. Growth of NWs occurred by initiating $AsH_3$ flow for a V/III ratio of 2.0, waiting 30 seconds, then opening the Al and Ga shutter. The growth was performed with the Al/Ga beam flux ratio set to a two-dimensional film composition of $Al_{0.37}Ga_{0.63}As$ and growth rate of 1 μm/hr, as determined by earlier thin film calibration growths. The wires were p-doped with Be concentration of $10^{18}$ $cm^{-3}$.

After growth, the morphology of the resulting NWs was observed by a JEOL JSM-7000F scanning electron microscope (SEM) in the secondary electron mode. NWs were removed from the substrate for further analysis in transmission electron microscopy (TEM) by sonicating in methanol for 1 to 2 minutes. A small volume (~10 μL) of the methanol solution was placed onto a holey carbon TEM support grid. After methanol evaporation, NWs were found to be dispersed onto the grid as observed by a JEOL 2010F high-resolution TEM. Energy dispersive x-ray spectroscopy (EDS) in the TEM, provided compositional analysis of the NWs using a probe size less than 1 nm, and measuring the Al $K\alpha_1$, Ga $K\alpha_1$, As $K\alpha_1$, and Au $L\alpha_1$ x-ray transitions.

NWs in methanol solution were dispersed onto silicon substrates for micro-photoluminescence (μPL) measurements in a continuous flow helium cryostat at 10 K and at room temperature. PL excitation and collection occurred through a microscope objective with numerical aperture of 0.7, providing a spot diameter of about 1 μm. The excitation was provided by a HeNe laser at wavelength of 632 nm and power of 2.5 μW. PL was resolved by a 75 cm grating spectrometer, and detected by a liquid nitrogen cooled Si CCD camera.

Polarization dependent μPL was measured at low temperatures (10 K) by rotation of a linear polarizer placed before the spectrometer along the emission path of the NW. The polarization dependence of the spectrometer and detector were confirmed to be negligible. Hence, the measured polarization dependence was that of the NWs only.

Finally, NWs were suspended by sonication in phosphate buffered saline solution with 30% glycerol by volume, and dropped onto a glass microscope slide for fluorescent image measurements using a Leica TCS SP5 confocal microscope with a Planapo 63x/1.3 NA objective. Excitation was provided by a 488 nm argon ion laser and detection by a Hamamatsu photomultiplier tube.

Figure 1(a) and (b) show the SEM images of AlGaAs NWs as-grown on GaAs (111)B substrates and post-sonicated onto silicon substrates, respectively. The AlGaAs NWs were grown for the same duration, growth conditions, and similar Au preparation method, as reported previously for GaAs NWs.[6] In the latter case, GaAs NWs exhibited a rod-like morphology with an average diameter in the range of 50 to 100 nm and height of 1 to 2 microns. In comparison, AlGaAs NWs exhibited a more strongly tapered morphology with comparatively large diameters in the range of 80 to 200 nm (100 nm typical) near the NW base and about 20 to 50 nm near the NW top, and heights of about 0.5 to 1 micron. This difference between AlGaAs and GaAs NW morphology can be explained by the difference in Al and Ga adatom migration lengths. NW growth occurs by diffusion of adatoms from the base to the top of NWs.[6] As NW height increases, adatom diffusion to the tops of NWs becomes increasingly unlikely and, instead of axial growth, deposition occurs on the NW sidewalls. Due to the shorter adatom migration length of Al compared to Ga,[7] sidewall deposition is more pronounced and leads to a more tapered morphology, a wider base, and a shorter height for AlGaAs as compared to GaAs NWs.

TEM and EDS measurements were carried out to determine the overall composition and spatial distribution of elements in the NWs. A number of EDS point measurements were made along several NWs, such as that shown in Figure 2. These EDS measurements indicate that the Al composition gradually declines along the NW length (and Ga composition correspondingly increases), while the As composition remains constant. Point EDS measurements of a half dozen similar NWs indicated an As composition of 50 atomic percent (as expected for AlGaAs). The Al composition (x) was in the range of 0.26 to 0.39 at the centre of the NW bases with an average measurement of 0.36. The broad size distribution and spacing of the Au catalysts used in this work inevitably introduce some variability in composition, morphology, and elemental distribution in the NWs. The NW composition was somewhat lower than the nominal composition of x = 0.37 expected from the thin film calibration growths. Similar observations have been reported elsewhere for AlGaAs NWs on GaAs substrates.[8] This difference in composition between NWs and 2-D films can be ascribed to differences between VLS and 2-D film growth mechanisms. A lower composition is expected in NWs where growth is dependent upon the diffusive transport of adatoms along the NW sidewalls towards the Au-NW interface.[9]

Compared to the base of the NWs, the Al composition at the top of the NWs (just below the Au particle) was determined by EDS to be in the range of x = 0.20 to 0.23 for the half dozen NWs examined. As the NW height increases, the Al composition evidently declines (and Ga composition correspondingly increases). This can be attributed to the limited Al adatom diffusion length already described in the context of NW morphology. Again, as the NW height begins to exceed the adatom migration length, sidewall growth begins to dominate as compared to axial growth. This effect is greatest for adatoms with the shortest diffusion length, so that an accumulation of Al would be expected around the outside of the NW.  To confirm this supposition, radial EDS measurements were performed across the base regions of numerous NWs (perpendicular to the growth direction).  Figure 3 shows the TEM image and radial EDS linescan of one such NW.  Contrast lines, intersecting the NW (perpendicular to the growth direction) in the TEM image indicate the presence of stacking faults, commonly observed in (111)B oriented NWs.[10, 11]  These stacking faults will not be considered here.  The EDS linescan reveals the anticipated core-shell structure with higher Al concentration near the outside of the NW.  Point EDS measurements across a half dozen similar NWs revealed an Al composition (x) of about 0.36 to 0.47 at the EDS maxima.  This represented an increase in Al composition in the NW shell compared to the core in the range of $\Delta x = 0.023$ to 0.044, or 0.033 on average.  Using the known relationship between AlGaAs composition and bandgap energy,[12] the core-shell structure corresponds to a confinement energy of about 46 meV near the NW base.

μPL spectra were measured at room temperature (300 K) and at 10 K on NWs dispersed onto Si substrates.  The μPL system had insufficient resolution to clearly discern whether individual NWs were being probed.  However, Figure 1(b) illustrated that the sonication procedure produced NWs with separations greater than 1 μm on average, compared to the excitation spot diameter of 1 μm, indicating that only a single NW at a time was likely excited.  Strong polarization anisotropy measurements, discussed below, also suggested that emission likely originated from single NWs.  Typical spectra at low temperature (10 K) and at room temperature (300 K) are shown in Figure 4.  First, the low temperature spectra revealed a high energy peak at 1.94 eV.  We start by assuming that this peak may be assigned to unresolved free, donor-bound, and/or acceptor-bound exciton transitions.  According to the composition dependence of the bandgap and accounting for the peak energy shift with temperature for bulk AlGaAs,[12] this peak then corresponds to an Al composition of x = 0.30. This composition is within the range measured by EDS for the base of the NWs. A somewhat higher intensity peak is observed at 1.877 eV with a shoulder at 1.895 eV, about 45 to 63 meV below the exciton-related peaks. This energy difference is consistent with band-to-acceptor and/or donor-to-acceptor transitions.[13] Lower energy peaks at 1.858, 1.835, and 1.815 eV (shoulder peak) are probably phonon replicas of the higher energy transitions.[14] Quantum confinement effects are unlikely to be a significant contributing factor to the PL results since the diameter of the NWs (~ 100 nm) is greater than the exciton Bohr radius estimated at 18 nm.[15] We note that comparable peak assignments have been made in both AlGaAs thin films[16, 17] and NWs.[10]

The highest intensity peak was observed at 1.731 eV at 10 K, or 1.72 eV at room temperature. The room temperature peak, assuming a band-to-band transition, corresponds to an average $Al_xGa_{1-x}As$ composition of x = 0.20. This is within the range of composition measured by EDS for the top of our NWs. The relative intensity of the PL peak at 1.731 eV (corresponding to the top of the NWs) compared to 1.940 eV (bottom of the NWs) suggests that excitons are diffusing and recombining preferentially in the potential minimum near the top of our NWs. One-dimensional exciton diffusion in excess of several microns has been observed in GaAs quantum wires.[18] In our case, exciton diffusion is likely driven by the potential gradient due to the Al distribution seen in Figures 2 and 3. The full width at half maximum (FWHM) of the µPL spectra was 62 meV at 10 K, and 120 meV at 300 K, significantly greater compared to the expected thermal broadening. These large peak widths are probably due to the compositional inhomogeneities within the NWs.

To further confirm that the µPL spectra were obtained from single NWs, the polarization dependence of the peak PL intensity at 10 K was measured. Figure 5a and b are polarization measurements performed by rotation of a linear polarizer placed before the spectrometer along the emission path of the NW. Fig. 5a shows the minimum ($I_{min}$) and maximum ($I_{max}$) PL emission determined by rotation of the polarizer where a change by a factor of 10 in peak intensity was observed. The degree of emission polarization, defined as $P = (I_{max} - I_{min}) / (I_{max} + I_{min})$ at the PL peak position, was 72% for our NWs. This strong emission polarization anisotropy is caused primarily by the dielectric mismatch between the NW and its surroundings, which causes strong suppression of the component of the electric field inside and perpendicular to the NW. Ruda and Shik[19] calculated that the emission from GaAs NWs should have a degree of polarization of approximately 92%. Although the polarization anisotropy of 72% for our AlGaAs NWs is lower than that for GaAs NWs, it is similar to that observed for other AlGaAs-GaAs core-shell heterostructures[20]. The measured anisotropy may depend upon the morphology and compositional gradients in the NWs.

To further confirm the observation of visible light emission from our NWs, a Leica confocal microscope was used to image the light emission from NWs suspended in phosphate buffered saline solution with 30% glycerol by volume. A prism spectrometer selecting a wavelength range of 670 to 730 nm, placed in front of the photomultiplier tube detector, allowed only the room temperature peak PL emission seen in Figure 4 to be detected. Figure 6 reveals the microscopy image (gray scale) of the NWs where the PL emission shown in red is superimposed. The image reveals a single NW on the far left that is about 1 μm in length and exhibiting a bright contrast on one end (lower side of the particle on the left), probably indicating the Au tip. The PL emission (shown in red) is superimposed on the microscopy image, revealing that the PL appears to emanate from the end of the wire near the Au particle. A series of such images were captured every 0.537 seconds, and the NW movement was tracked using ImageJ analysis software.[21] Figure 7 shows the mean square displacement (MSD), $<r^2>$, versus time $t$ for two NWs in the solution. Considering the diffusion equation $<r^2> = 4Dt$, a linear fit to the data (solid line in Figure 7) gives a diffusion coefficient of $D = 0.4$ μm$^2$/s. To verify that the PL emission in Figure 6 is from a single NW, the diffusion coefficient may be calculated from the hydrodynamic stick theory for rods[22, 23]:

$$D = \frac{k_B T}{3\pi\eta L}\ln(L/d) \qquad\qquad (1)$$

where $L$ is the NW length, $d$ is the diameter, $k_B$ is Boltzmann constant, $T$ is the temperature (300K), and $\eta$ is the viscosity of the solution. For a typical NW in our experiment, $L = 800$ nm and $d = 100$ nm on average, and $\eta = 2.26$ cP for phosphate buffered saline solution with 30% glycerol by volume.[24] This yields $D = 0.5$ μm$^2$/s, in good agreement with the experimental result. These results confirm the visible light emission from a single NW can be detected, and suggest possible applications in biological imaging.

In conclusion, Be-doped Al$_x$Ga$_{1-x}$As (x = 0.37) NWs were grown on GaAs (111)B substrates using a gas source molecular beam epitaxy system. Strong photoluminescence polarization measured at 10 K was observed, indicative of emission from single NWs. Visible light emission from the NWs at room temperature was confirmed by confocal microscopy. EDS and μPL measurements indicated an Al composition gradient and a core-shell structure, which drove exciton diffusion to the top of NWs where recombination occurred. The core-shell structure was self-directed during growth, evolving as a consequence of the different Al and Ga adatom diffusion lengths.

**Acknowledgements:** This work was supported by a Nano Innovation grant, the Ontario Photonics Consortium, and the Canadian Foundation for Innovation. The authors thank Brad

Robinson and the staff of the CEDT for the GS-MBE growths and the enlightening discussions, and Fred Pearson for assistance with the TEM.

**References**

1. A.P. Alivisatos, Science 271 (1996) 933.

2. C.M. Lieber, Solid State Commun. 107 (1998) 607.

3. R.E. Smalley and B.I. Yakobson, Solid State Commun. 107 (1998) 597.

4. C.M. Lieber and Z.L. Wang, MRS Bulletin 32 (2007) 99.

5. Fu, W. Gu, B. Boussert, K. Koski, D. Gerion, L. Manna, M.L. Gros, C.A. Larabell, and A.P. Alivisatos, Nano Lett. 7 (2007) 179.

6. M.C. Plante and R.R. LaPierre, J. Cryst. Growth 286 (2006) 394.

7. T. Shitara, J.H. Neave, and B.A. Joyce, Appl. Phys. Lett. 62 (1993) 1658.

8. Z.H. Wu, M.Sun, X.Y. Mei, and H.E. Ruda, Appl. Phys. Lett. 85 (2004) 657.

9. J. Johansson, L. S. Karlsson, C. P. T. Svensson, T. Martensson, B. A. Wacaser, K. Deppert, L. Samuelson and W. Seifert, Nat. Mater. 5, 574 (2006).

10. S. Bhunia, T. Kawamura, S. Fujikawa, H. Nakashima, K. Furukawa, K. Torimitsu and Y. Watanabe, Thin Solid Films 464-465, 244 (2004).

11. L.E. Jensen, M.T. Bjork, S. Jeppesen, A.I. Persson, B.J. Ohlsson, and L. Samuelson, Nano Lett. 4, 1961 (2004).

12. Vurgaftman and J.R. Meyer, J. Appl. Phys. 89 (2001) 5815.

13. V. Swaminathan, J.L. Zilko, W.T. Tsang and W.R. Wagner, J. Appl. Phys. 53 (1982) 5163.

14. S.J. Chua, S.J. Xu, and X.H. Tang, Sol. State. Commun. 98 (1996) 1053.

15. S. Adachi, J. Appl. Phys. 58 (1985) R1.

16. Monemar, K.K. Shih, and G.D. Pettit, J. Appl. Phys. 47 (1976) 2604.

17. G. Oelgart, R. Schwabe, M. Heider, and B. Jacobs, Semicond. Sci. Technol. 2 (1987) 468.

18. Y. Nagamune, H. Watanabe, F. Sogawa, and Y. Arakawa, Appl. Phys. Lett. 67 (1995) 1535.

19. H.E. Ruda and A. Shik, Phys. Rev. B 72 (2005) 115308.

20. L.V. Titova, T.B. Hoang, H.E. Jackson, L.M. Smith, J.M. Yarrison-Rice, Y. Kim, H. J. Joyce, H. H. Tan, and C. Jagadish, Appl. Phys. Lett. 89 (2006) 173126.

21. http://rsb.info.nih.gov/ij/

22. R. Vasanthi, S. Bhattacharyya, and B. Bagchi, J. Chem. Phys. 116 (2002) 1092.

23. James M. Tsay, Sören Doose, and Shimon Weiss, J. Am. Chem. Soc. 128 (2006) 1639.

24. K. Luby-Phelps, S. Mujumdar, R. B. Mujumdar, L. A. Ernst, W. Galbraith, and A. S. Waggoner, Biophys. J. 65 (1993) 236.

**Figure Captions**

Figure 1. (a) Tilted view (30˚) SEM image of as-grown AlGaAs NWs. (b) Plan view SEM image of NWs sonicated onto a Si substrate.  Scale bars are 1 μm.

Figure 2. TEM image and EDS point measurements along the centre axis of an AlGaAs NW. The Au nanoparticle is visible at the top of the NW on the right-hand side.

Figure 3. Dark field TEM image and EDS linescan across an AlGaAs NW (perpendicular to the growth direction).  Scale bar is 20 nm.

Figure 4.   μPL spectra at 10 K and at room temperature (300 K) for an AlGaAs NW. The 10 K spectra above 1.80 eV and the room temperature spectra are multiplied by 5 and 10, respectively, for clarity.

Figure 5.  Polarization dependence of μPL emission at 10 K.  0 and 90˚ indicate the relative angle of the linear polarizer.

Figure 6.  PL image (red) superimposed on the differential interference contrast image (grayscale).  Scale bar is 3 μm.

Figure 7.  Mean square displacement (MSD) versus time for AlGaAs NWs dispersed in phosphate buffered saline solution with 30% glycerol by volume.  Data points indicate measurements for two different NWs.  The solid line is a linear fit to the data.

The last page is the Table of Contents Graphic

Figure1

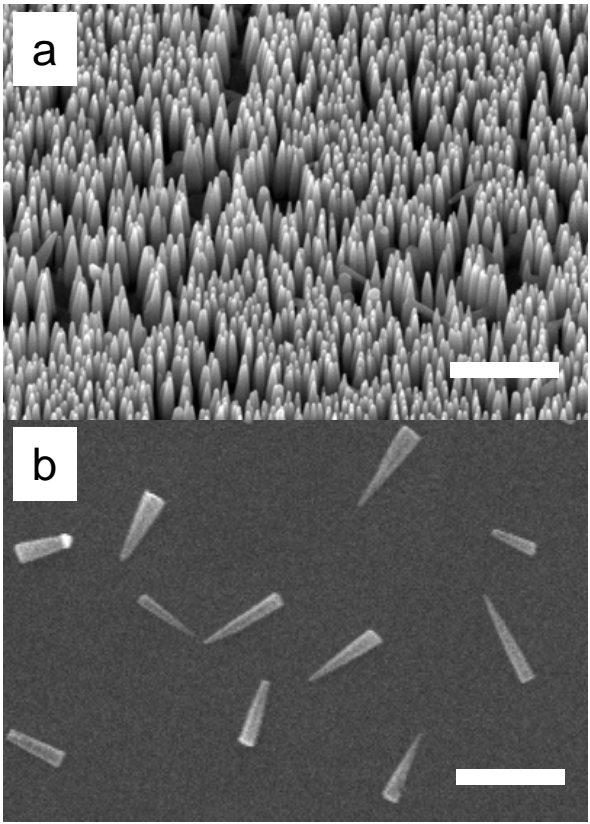

Figure 2

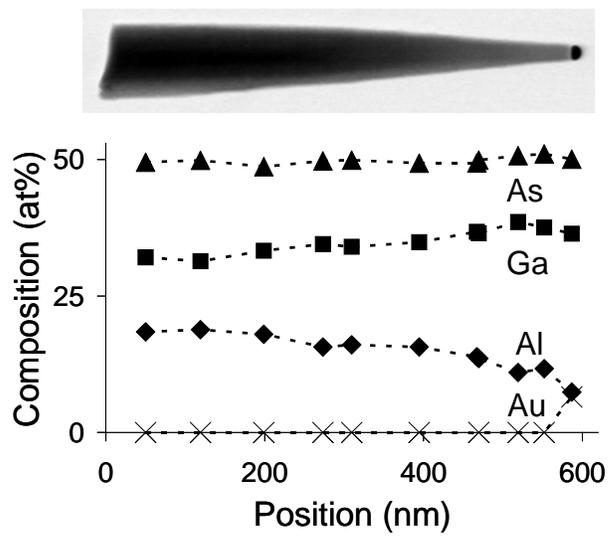

Figure 3

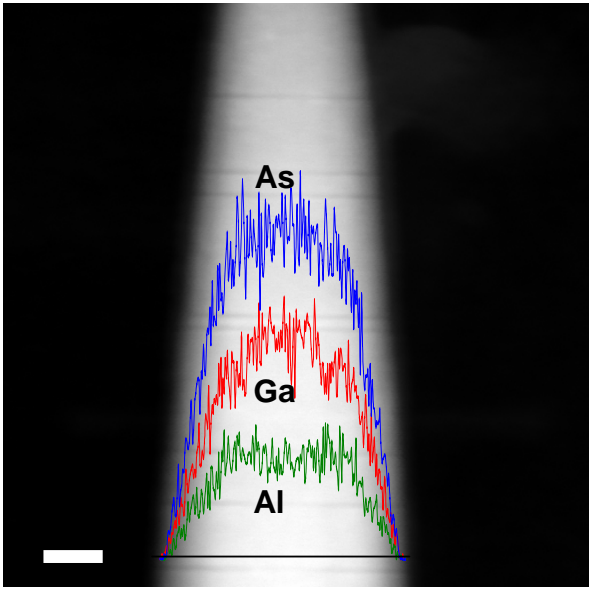

Figure 4

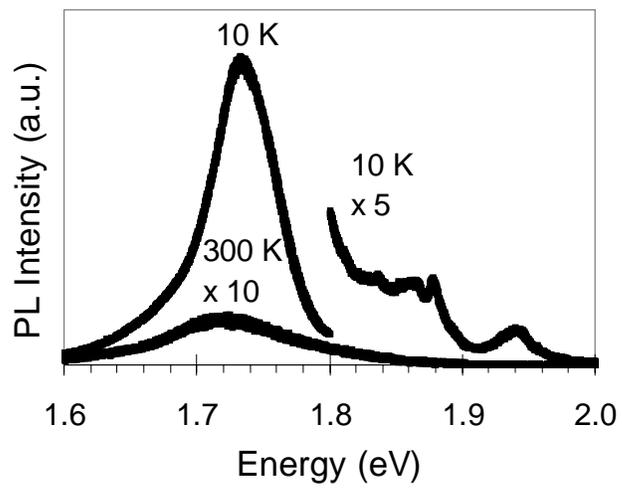

Figure 5

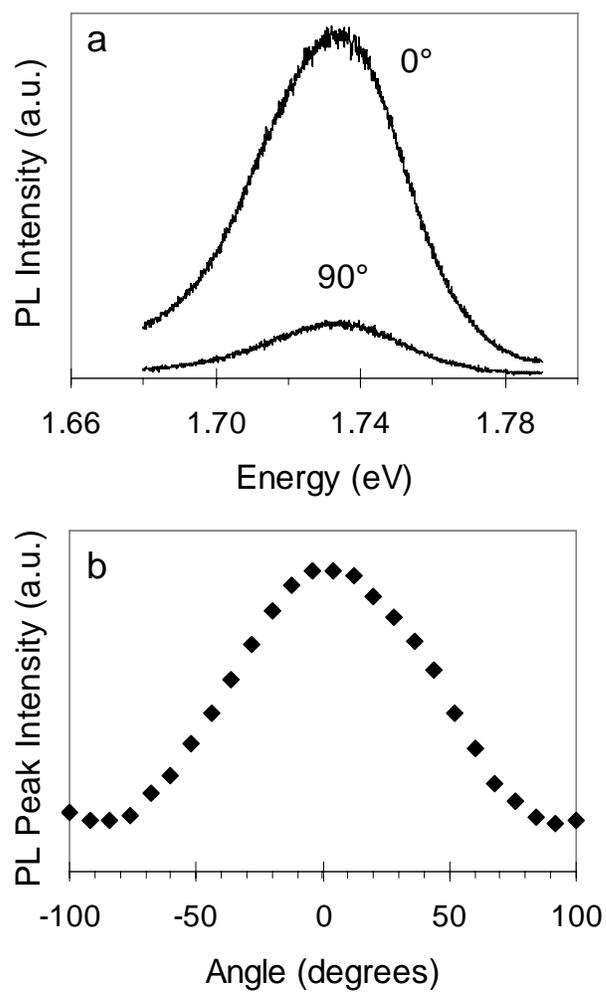

Figure 6

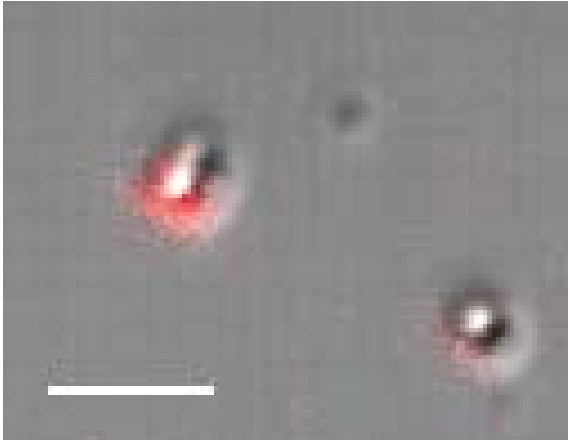

Figure 7

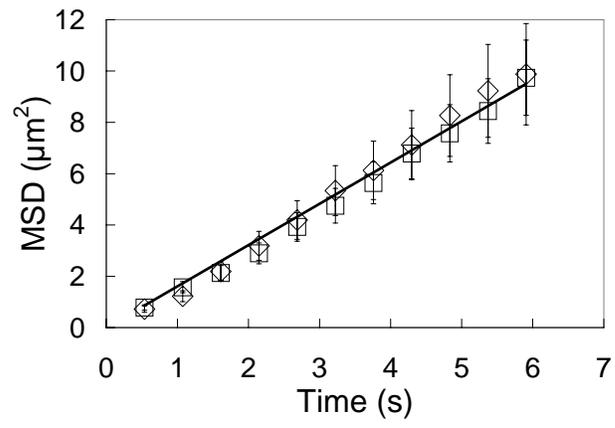

Table of contents graphic

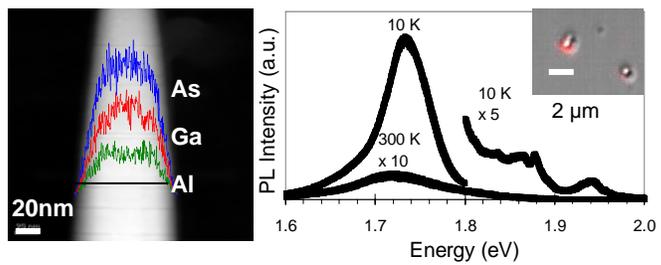